\documentclass[aps,prd,twocolumn,nofootinbib,superscriptaddress,showpacs,floatfix,preprintnumbers]{revtex4-2}

\usepackage[dvipsnames]{xcolor}     
\definecolor{lcolor}{rgb}{0.5,0,0}
\definecolor{citcolor}{rgb}{0,0,1}
\usepackage[breaklinks,colorlinks,urlcolor=blue,citecolor=citcolor,linkcolor=lcolor,linktoc=all]{hyperref}
\usepackage{color}
\usepackage{graphicx}	
\graphicspath{{./figures/}}
\usepackage[utf8]{inputenc}
\usepackage{amsmath} \usepackage{amssymb}
\usepackage[dvipsnames]{xcolor}      
\usepackage{bm,bbm,bbold}
\usepackage{booktabs}
\usepackage{dcolumn}
\usepackage{comment}

\usepackage{tikz}
\usepackage[customcolors]{hf-tikz}
\usepackage{mciteplus}

\begin{document}

\title{Symmetry-based resolution of pseudogauge ambiguities in local equilibrium}

\author{Carlos Hoyos}
\email{hoyoscarlos@uniovi.es}
\affiliation{Departamento de F\'{\i}sica and Instituto de Ciencias y Tecnolog\'{\i}as Espaciales de Asturias (ICTEA), Universidad de Oviedo,c/ Leopoldo Calvo Sotelo 18, ES-33007, Oviedo, Spain}

\begin{abstract}
The total angular momentum current can be decomposed into orbital and spin contributions in different ways, known as pseudogauges. This freedom leads to ambiguities in the definition of local-equilibrium density operators, which in turn affect estimates of spin polarization in heavy-ion collisions. In this work, the pseudogauge ambiguity, together with other ambiguities associated with improvements of conserved currents, is reformulated in terms of spurious symmetries corresponding to conserved currents with vanishing total charge. A prescription for the unambiguous definition of a local-equilibrium density operator, defined as deviating from global equilibrium only by small gradients, is introduced using the currents associated with genuine symmetries. The resulting density operator is invariant under transformations that add improvement terms to local currents, including the energy-momentum tensor.
\end{abstract}

\maketitle

\section{Introduction}

The total angular momentum current in relativistic quantum field theory admits different decompositions into orbital and spin components. This nonuniqueness becomes particularly relevant in the context of spin polarization phenomena in relativistic heavy-ion collisions, where it is commonly referred to as the pseudogauge freedom of the energy--momentum and spin tensors \cite{Becattini:2018duy,Li:2020eon,Fukushima:2020ucl,Buzzegoli:2021wlg,Drogosz:2024rbd}. Similar issues also arise in other contexts, such as the decomposition of the nucleon spin \cite{Leader:2013jra}. For reviews on spin hydrodynamics and polarization in relativistic fluids, see \cite{Florkowski:2018fap,Speranza:2020ilk,Florkowski:2019qdp}.

Experimental data from relativistic heavy-ion collisions indicate that the produced matter is well described, over a broad range of observables, by relativistic viscous hydrodynamics \cite{Heinz:2013th,Gale:2013da,Teaney:2009qa}. This supports the picture that the system rapidly approaches a regime close to local thermodynamic equilibrium at early times after the collision. While modern developments suggest that hydrodynamic behavior may emerge before complete local thermalization is achieved, a process often referred to as hydrodynamization \cite{Romatschke:2019ejr,Florkowski:2017olj}, local equilibrium remains the natural leading-order description of the quark--gluon plasma.

In this framework, the local equilibrium state is expected to be independent of the choice of pseudogauge. However, in practical constructions based on local equilibrium density operators, a residual pseudogauge dependence generally appears, which can have a quantitatively relevant impact on spin-dependent observables \cite{Buzzegoli:2021wlg}. This issue was recently addressed in \cite{Becattini:2025twu}, where a pseudogauge-invariant local equilibrium density operator was constructed by combining the energy--momentum tensor and spin tensor with coefficients depending on derivatives of the thermal four-velocity. Nevertheless, a residual ambiguity remains associated with the possibility of adding divergenceless contributions to the spin tensor. Similar ambiguities arise for conserved currents of global symmetries, where identically conserved (improvement) terms can be added without affecting global charges. 

In this work, we clarify the origin of these ambiguities by showing that the transformations by improvements of local currents can be reinterpreted as an invariance of the density operator under the addition of terms proportional to spurious charges with constant conjugate variables. We then construct a generalization to coordinate-dependent variables such that the density operator remains invariant under the full class of improvement transformations. Invariance under pseudogauge transformations can be derived as a consequence of this.

We will define a local equilibrium state as one that at any point deviates from global equilibrium by contributions that can be expanded in a small gradient expansion. There could be other interesting nonequilibrium states where spin relaxes more slowly than momentum and one deviates from this situation, but we will not address those in this work. We propose that the local thermal equilibrium density operator should depend only on intensive quantities conjugate to charges of genuine symmetries. We further propose a systematic way to separate genuine conserved currents, associated with physical symmetries, from identically conserved spurious currents with trivial symmetry generators. With this proposal, we are able to show that the transformations of the density operator imply the resulting operator is equivalent to the one constructed in \cite{Becattini:2025twu}. This leads to a formulation free of ambiguities related to the choice of energy--momentum tensor, spin current, or other conserved currents.

\section{Constant velocity and chemical potential}

Global thermal equilibrium is a mixed state characterized uniquely by the conserved charges of the global symmetries of the system. In the grand canonical ensemble the density operator $\widehat\rho$ depends only on the intensive variables that are conjugate to the charges. For simplicity we will consider the case of a single Abelian internal symmetry. The intensive variables are the constant thermal velocity $\beta_\mu$, conjugate to energy and momentum, and the reduced chemical potential $\zeta$, conjugate to the charge of the internal symmetry. From the extremization of the von Neumann entropy $S=-\operatorname{tr}(\widehat \rho \log\widehat \rho)$ by the density operator one can deduce its form \cite{Zubarev:1974,Zubarev:1996,Becattini:2012tc,Hongo:2016mqg}
\begin{equation}\label{eq:globaleq}
    \widehat \rho=\frac{e^{-\widehat W}}{Z},\quad \, \widehat W=\beta_\mu\widehat P^\mu+\zeta \widehat Q\,,
\end{equation}
where $Z=\operatorname{tr}\left(e^{-\widehat W}\right)$ and $\widehat P^\mu$, $\widehat Q$ are the energy and momentum, and Abelian charge operators respectively. 

In principle the density operator can be expressed in terms of local operators, the energy-momentum tensor $\widehat T^{\mu\nu}(x)$, and the Abelian current $\widehat J^\mu(x)$
\begin{equation}\label{eq:charges}
    \widehat P^\mu=\int d^d x \,\widehat T^{0\mu}\,,\quad \widehat Q=\int d^d x\, \widehat J^0\,.
\end{equation}
We have chosen constant time slices for the volume integrals for clarity of the presentation, but the analysis can be extended to other spatial hypersurfaces. In particular, thermalization in heavy ion collisions is expected to take place at constant Milne time.

There is an ambiguity in writing the density operator using local current operators. The charge operators do not change when the currents are modified by improvement terms\footnote{If there are boundaries, the charge operators are still invariant if one adds appropriate boundary terms depending on the potential operators.}
\begin{subequations}\label{eq:improvements}
\begin{eqnarray}
    \widehat T^{\mu\nu}_{\rm imp} &=& \widehat T^{\mu\nu}+\frac{1}{2}\partial_\lambda\left( \widehat \Phi^{\lambda,\mu\nu}
+\widehat \Phi^{\mu,\nu\lambda}+ \widehat \Phi^{\nu,\mu\lambda}\right)\,,\\
\widehat J^\mu_{\rm imp} &=&  \widehat J^\mu+\partial_\lambda \widehat M^{\lambda\mu}\,.
\end{eqnarray}
\end{subequations}
provided the operators acting as ``potentials'' of the transformation satisfy
\begin{equation}\label{eq:pseudocond}
\widehat \Phi^{\mu,\alpha\beta}=-\widehat \Phi^{\mu,\beta\alpha}\,,\quad  
\widehat M^{\lambda\mu}=-\widehat M^{\mu\lambda}\,.
\end{equation}
However, if these operators exist, there are additional conserved current operators that one can define
\begin{subequations}\label{eq:spuriouscurrents}
\begin{eqnarray}
\label{eq:pseudogaugeSpin}  \widehat  J^{\mu\nu}_\Phi &=& \frac{1}{2}\partial_\lambda\left( \widehat \Phi^{\lambda,\mu\nu}
+\widehat \Phi^{\mu,\nu\lambda}+ \widehat\Phi^{\nu,\mu\lambda}\right),\\
 \widehat J_M^\mu &=&\partial_\lambda \widehat M^{\lambda\mu}\,.
\end{eqnarray}
\end{subequations}
By construction, these additional currents are conserved off-shell and their densities are total spatial derivatives. In general, this implies that the total charge vanishes, but there can be some exceptions if they are topological currents, like the monopole current of an Abelian gauge field in $2+1$ dimensions, where $\widehat M^{\lambda\mu}=\epsilon^{\lambda\mu\nu}\widehat A_\nu$ is not a gauge-invariant operator. Barring topological currents, any other current of the form
\begin{equation}
    \widehat J^{\mu I} =\partial_\lambda \widehat M^{\lambda\mu I},\quad \widehat M^{\lambda\mu I}=-\widehat M^{\mu\lambda I}\,, 
\end{equation}
with $I$ internal or spacetime indices, has a vanishing total charge
\begin{equation}
    \widehat Q^I=\int d^{d} x \, \widehat J^{0 I}=\int d^{d} x \, \partial_i\widehat M^{i0 I}=0\,.
\end{equation}
Currents of this type correspond to spurious symmetries, in the sense that the unitary operators of symmetry transformations, given by the exponential of the total charge, $\widehat U=e^{i\alpha_I \widehat Q^I}$, equal the identity. We will refer to the currents for these spurious symmetries as ``spurious currents.'' Similarly, currents for genuine symmetries will be denoted as ``genuine currents.''

This gives us an alternative way of thinking about the transformations by improvements \eqref{eq:improvements}. The transformed currents are linear combinations of conserved currents of genuine and spurious symmetries. From this perspective, the transformation can be understood as a transformation of the density operator rather than of the local current operators. The density operator changes to $\widehat {\widetilde W}=\widehat W+\Delta \widehat W$ with 
\begin{equation}
   \Delta \widehat W=\beta_\mu \int d^d x \widehat J_\Phi^{0\mu}+\zeta \int d^d x \widehat J_M^0\,.
\end{equation}
The new operator has the same form as a density operator with nonzero chemical potentials for the spurious currents. Since the total charges of the spurious currents vanish, the transformed operator is actually equal to the original operator. But we can in fact generalize this transformation and allow for arbitrary chemical potentials for the spurious currents, with values independent of the thermal velocity and chemical potential of the Abelian current. Therefore, the transformation effectively acting as an improvement can be seen as part of a larger redefinition of global equilibrium density operators, where one can introduce arbitrary constant chemical potentials for any number of spurious currents
\begin{equation}\label{eq:Wtransf}
    \Delta \widehat W=\sum_I \zeta_I \widehat Q^I\,.
\end{equation}
Although the local form of the currents in the definition of the density operator is modified, all these operators are equivalent to each other since the total charge of the spurious symmetries vanishes $ \widehat Q^I=0$.

\section{Coordinate-dependent intensive variables}

The extension from global to local thermal equilibrium requires promoting intensive variables to slowly varying functions of space and time $x=(t,\bm{x})$. If the spin current $\widehat S^{\mu,\alpha\beta}$ is nonzero it is customary to define the density operator as\footnote{We use the conventions $\partial_{[\alpha}\beta_{\beta]} =\frac{1}{2}(\partial_\alpha \beta_\beta-\partial_\beta \beta_\alpha)$, and $\partial_{(\alpha}\beta_{\beta)} =\frac{1}{2}(\partial_\alpha \beta_\beta+\partial_\beta \beta_\alpha)$.}
\begin{equation}\label{eq:primaryloceq}
    \widehat W(\beta_\mu,\zeta)=\int d^d x \, \left(\beta_\mu \widehat T^{0\mu}+\frac{1}{2} \partial_{[\alpha}\beta_{\beta]} \widehat S^{0,\alpha\beta}+\zeta \widehat J^0\right)\,.
\end{equation}

A particular case is when $\zeta$ is constant and $\beta_0=\bar\beta_0$, $\beta_i=\bar \beta_i-\Omega_{ij}x^j$, with $\bar\beta$, $\Omega_{ij}$ constant. In this case there is a constant vorticity which acts as a chemical potential for the angular momentum. The term proportional to the spin current in \eqref{eq:primaryloceq} guarantees that the density operator only depends on the conserved total angular momentum. Since the angular momentum is conserved, this is also considered to be a global equilibrium state.\footnote{Another way to arrive at the same expression is to introduce a constant thermal velocity and a constant chemical potential for the total angular momentum, and then combine the orbital part of the angular momentum with the velocity term.}  However, since it involves a velocity that depends on the spatial coordinates, we treat it here together with the local equilibrium states.  Note that the second term in \eqref{eq:primaryloceq} resembles a chemical potential for the spin current proportional to the thermal vorticity. In \cite{Becattini:2018duy}, the possibility of having an arbitrary ``spin chemical potential'' was also considered. We will see this in more detail later, when we discuss pseudogauge transformations.

When the intensive variables are constant,  having a combination of genuine and spurious currents makes no difference, but this is no longer the case when the intensive variables depend on the coordinates. When the density operator with nonzero chemical potentials for spurious currents \eqref{eq:Wtransf} is extended in this way we encounter a potential problem: terms of the form
\begin{equation}\label{eq:spuriousleq}
  \Delta \widehat W=  \int d^d x \, \zeta_I(x)\widehat J^{0 I}(x)\,,
\end{equation}
are no longer total derivatives and the density operator would become dependent on the generalized transformations. 
This can be amended by modifying the na\"{\i}ve extension to coordinate-dependent intensive variables \eqref{eq:spuriousleq} by 
\begin{equation}\label{eq:Wtransf2}
    \Delta \widehat W= \int d^d x \left(\zeta_I \widehat J^{0 I}+\partial_i \zeta_I \widehat M^{i0 I}\right)=\int d^d x\,\partial_i\left(\zeta_I \widehat M^{i0 I}\right)\,.
\end{equation}
By definition of a spurious current this is always possible. In this way, the family of generalized transformations that keep the density operator invariant extends to nonzero vorticity and local equilibrium.  One still has the issue of correctly identifying the current operators in \eqref{eq:primaryloceq} that determine the local thermal equilibrium density operator, which we will discuss below.

\section{Density operator for local thermal equilibrium}

We have established that any density operator of the form given in \eqref{eq:primaryloceq} is invariant under the generalized improvement transformations. Therefore, each choice of a current in \eqref{eq:primaryloceq} corresponds to a physically distinct state. If the current is a combination of a genuine and a spurious current, the density operator will depend on the chemical potential conjugate to the spurious current, which is being forced to be proportional to the thermal velocity. We propose that the local thermal equilibrium state should depend only on intensive variables conjugate to genuine currents. Assuming this is correct, the problem of constructing the local thermal equilibrium density operator will reduce to that of identifying which operators correspond to genuine currents.

We will now provide a concrete proposal distinguishing the conserved current of a genuine symmetry from any other combination with spurious currents.

Conformal field theories are a good starting point, since genuine symmetries have associated currents that are quasiprimary operators, while spurious currents are descendants. 

Given the generators of dilatations, $\widehat  D$, and special conformal transformations, $\widehat K_\mu$, a local quasiprimary operator of conformal dimension $\Delta$ in a CFT satisfies \cite{DiFrancesco:1997nk}
\begin{equation}
    i\left[\widehat D,\widehat O_\Delta(0)\right]=\Delta \,\widehat O_\Delta(0),\ \ i\left[\widehat K_\mu,\widehat O_\Delta(0)\right]=0\,.
\end{equation}
 Our proposal is that genuine currents are operators satisfying these conditions. Spurious currents on the other hand will all be descendant operators, i.e., operators that can be obtained from commutators with the generators of translations $\widehat J^{\mu I}=i\left[\widehat P_\lambda,\widehat M^{\lambda\mu I}\right]$. Note that this automatically selects the energy-momentum tensor as the symmetric and traceless operator, and correctly identifies topological currents as quasiprimaries (see, e.g.,~\cite{Borokhov:2002ib}) and thus corresponding to genuine symmetries.

This split between ``quasiprimary'' and ``descendant'' currents is also useful in nonconformal theories, whenever they are obtained from a conformal theory by deformations that introduce an explicit or anomalous breaking of conformal symmetry. Clearly the Standard Model, and QCD within it, fall into this category, so this covers most cases of interest. In the case of explicit breaking, the Lagrangian density of the theory is a deformation of a conformal theory of the form
\begin{equation}\label{eq:conformalbreak}
\widehat {\cal L}= \widehat {\cal L}_{CFT}-\sum_A g_A \widehat O^A\,,
\end{equation}
where in $d+1$ spacetime dimensions $\widehat O^A$ are relevant operators of conformal dimension $\Delta_A<d+1$ and $g_A$ are the couplings. 

Global symmetries, other than chiral symmetries, are usually not directly affected by breaking of conformal invariance, so the ``quasiprimary'' currents are the same as in the associated conformal theory. The energy-momentum tensor on the other hand is affected; in particular, it is no longer traceless. In this case the continuation from the quasiprimary operator of the conformal theory is the symmetric energy-momentum tensor operator satisfying the Ward identity for the broken conformal invariance
\begin{equation}\label{eq:Tmm}
    \widehat T^\mu_{\ \mu}=\sum_A \beta_A \widehat O^A\,,
\end{equation}
where $\beta_A$ are the beta functions of the couplings breaking conformal invariance. For an explicit breaking like \eqref{eq:conformalbreak} $\beta_A=(d+1-\Delta_A)g_A$. Note that the spin current operator for a symmetric energy-momentum tensor would be conserved. However, even if it is nonvanishing it cannot be a genuine current. In ordinary relativistic quantum field theories, the only symmetries of the theory are Poincaré invariance and internal global symmetries, with possibly conformal and/or supersymmetric extensions. If there were nontrivial conserved spin charge operators, one would be able to construct nontrivial symmetry generators associated with these charges. Since these symmetries do not exist, there cannot be nontrivial spin charge operators. It is consistent in this case not to include the would-be spin current in the total angular momentum, since its contribution would be vanishing.

As concrete examples consider the free massive fields of a real scalar $\widehat \phi$ and a Dirac fermion $\widehat\psi$. Their Lagrangians are deformations \eqref{eq:conformalbreak} of the conformal invariant massless theory with the operators
\begin{subequations}
\begin{eqnarray}
    \widehat O_s &=&\frac{1}{2}{\widehat\phi}^2, \ g_s=m_s^2,\ \beta_s=2m_s^2,\\
    \widehat O_f &=&\widehat{\overline{\psi}}\widehat\psi,\ g_f=m_f,\ \beta_f=m_f\,.
\end{eqnarray}
\end{subequations}
The symmetric energy-momentum tensors satisfying \eqref{eq:Tmm} are\footnote{We introduce a  mostly minus Minkowski metric $\eta_{\mu\nu}$, the corresponding gamma matrices $\gamma^\mu$, and derivatives acting on both sides $\overset{\leftrightarrow}{\partial}_\mu=\overset{\rightarrow}{\partial}_\mu-\overset{\leftarrow}{\partial}_\mu$.}
\begin{subequations}
    \begin{eqnarray}
\notag        \widehat T^{\mu\nu}_{\rm s}&=&: \partial^\mu \widehat \phi\, \partial^\nu\widehat\phi-\frac{1}{2}\eta^{\mu\nu}\left(\partial_\lambda \widehat\phi \partial^\lambda \widehat \phi-m_s^2\widehat\phi^2\right)\\
        \notag & &-\frac{d-1}{4d}\left(\partial^\mu\partial^\nu-\eta^{\mu\nu}\partial_\lambda\partial^\lambda \right)\widehat \phi^2 :\,,\\
\notag        \widehat T^{\mu\nu}_{\rm f}&=&:\frac{i}{4}\widehat{\overline{\psi}}\left(\gamma^\mu\overset{\leftrightarrow}{\partial^\nu}+\gamma^\nu \overset{\leftrightarrow}{\partial^\mu}\right)\widehat\psi:\,.
    \end{eqnarray}
\end{subequations}
The addition of the last term to the scalar energy-momentum tensor is further supported by considerations about the finiteness of matrix elements of the energy-momentum tensor in the renormalized theory \cite{Callan:1970ze}. This term can be obtained from a variation of the action with respect to a background metric if there is a nonminimal coupling to the Ricci scalar curvature $\Delta \widehat {\cal L}=-\frac{1}{2} \xi_d R\widehat \phi^2$, with  $\xi_d=\frac{d-1}{4d}$. For a minimal or general nonminimal coupling, the genuine energy-momentum tensor differs from the one obtained through the variation. 

\section{Pseudogauge transformations}

We will explain now how pseudogauge transformations emerge from the generalized transformations \eqref{eq:Wtransf2}, and show that this formulation implies pseudogauge invariance of the density operators.

Let us discuss first the case of global equilibrium with nonzero vorticity in more detail. The density operator takes the form
\begin{equation}\label{eq:equilvort}
    \widehat W= \bar \beta_\mu \widehat P^\mu+\frac{1}{2}\Omega_{ij}\widehat {\cal J}^{ij}+\zeta \widehat Q\,,
\end{equation}
 This form is equivalent to \eqref{eq:primaryloceq}, after separating the terms proportional to the vorticity.
Using the genuine (symmetric) energy-momentum tensor and Abelian current, the energy, momentum and charge have the form in \eqref{eq:charges}, and the angular momentum operator is
\begin{equation}
    \widehat {\cal J}^{ij}=\int d^dx\left( x^i \widehat T^{0j}-x^j \widehat T^{0i}\right)\,.
\end{equation}

We can use the transformations \eqref{eq:Wtransf2} to write the density operator in an equivalent form, but with an energy-momentum tensor that is not necessarily symmetric and with an angular momentum operator that depends on the associated spin current operator $\widehat S^{\mu,\alpha\beta}$, for instance both could correspond to the canonical operators obtained from the Lagrangian using Noether's procedure.

The transformation uses the spurious current \eqref{eq:pseudogaugeSpin}, fixing $\widehat \Phi^{\mu,\alpha\beta}=-\widehat S^{\mu,\alpha\beta}$ and $\zeta_\mu=\beta_\mu=\bar\beta_\mu-\Omega_{ij}x^j$. After applying it, the density operator has the form in \eqref{eq:equilvort}, with energy and momentum operators given by \eqref{eq:charges}, but with the energy-momentum tensor operator replaced by  
\begin{equation}
   \widehat  T^{\mu\nu}\longrightarrow \widehat T^{\mu\nu}_{\rm as}=\widehat T^{\mu\nu}-\frac{1}{2}\partial_\lambda\left( \widehat S^{\lambda,\mu\nu}
+\widehat S^{\mu,\nu\lambda}+ \widehat S^{\nu,\mu\lambda}\right)\,.
\end{equation}

The angular momentum operator after the transformation becomes
\begin{equation}
    \widehat {\cal J}_{\rm as}^{ij}=\int d^dx\left( x^i \widehat T_{\rm as}^{0j}-x^j \widehat T_{\rm as}^{0i}+\widehat S^{0,ij}\right)\,.
\end{equation}
Further transformations \eqref{eq:Wtransf2} using the spurious currents \eqref{eq:pseudogaugeSpin} with $\zeta_\mu=\beta_\mu$ effectively act as pseudogauge transformations
\begin{subequations}\label{eq:pseudogauge}
\begin{eqnarray}
    \widetilde{\widehat T^{\mu\nu}} &=& \widehat T_{\rm as}^{\mu\nu}+\frac{1}{2}\partial_\lambda\left( \widehat \Phi^{\lambda,\mu\nu}
+\widehat \Phi^{\mu,\nu\lambda}+ \widehat \Phi^{\nu,\mu\lambda}\right)\,,\\
\widetilde{\widehat S^{\mu,\alpha\beta}} &=& \widehat S^{\mu,\alpha\beta}-\widehat \Phi^{\mu,\alpha\beta}\,.
\end{eqnarray}
\end{subequations}
 Therefore, invariance under the transformations in \eqref{eq:Wtransf2} results in invariance under pseudogauge transformations. Note that this implies that all density operators of the form \eqref{eq:equilvort} are equivalent, for any choice of energy-momentum tensor and associated spin current, as expected.

At local equilibrium, the pseudogauge invariant operator constructed in \cite{Becattini:2025twu}  can be found in this approach following similar steps: the density operator \eqref{eq:primaryloceq} is defined using the genuine (symmetric) energy-momentum. Then, the density operator is transformed with \eqref{eq:Wtransf2}  using the spurious currents \eqref{eq:pseudogaugeSpin} with $\widehat \Phi^{\mu,\alpha\beta}=-\widehat S^{\mu,\alpha\beta}$ and $\zeta_\mu=\beta_\mu$
\begin{equation}\label{eq:DeltaWlocal}
   \Delta \widehat W= -\int d^d x \left( \beta_\mu\widehat J_{S}^{0\mu }+\frac{1}{2}\partial_i\beta_\mu\left(\widehat S^{i,0\mu}+\widehat S^{0,\mu i}+\widehat S^{\mu, 0i}\right)\right)\,.
\end{equation}
The resulting expression for the transformed density operator coincides with the one given in \cite{Becattini:2025twu}. 

Pseudogauge invariance can actually be extended in some form to other density operators. We start with the density operator
\begin{equation}\label{eq:W0}
    \widehat W_{\rm ref}=\int d^d x \, \left(\beta_\mu \widehat T_{\rm ref}^{0\mu}+\left(\frac{1}{2} \partial_{[\alpha}\beta_{\beta]}+\Omega^{(1)}_{\alpha\beta}\right) \widehat S_{\rm ref}^{0,\alpha\beta}+\zeta \widehat J^0\right)\,.
\end{equation}
Where, following \cite{Becattini:2018duy}, we have introduced a nonequilibrium contribution $\Omega^{(1)}_{\alpha\beta}$ in the spin current term. The nonequilibrium contribution must vanish at global equilibrium. It then follows that at local equilibrium it should be of higher order than the thermal vorticity in the gradient expansion.\footnote{An equivalent statement applies if the operator is written using the total angular momentum
\begin{equation}\label{eq:W0b}
    \widehat W_{\rm ref}=\int d^d x \, \left(\tilde \beta_\mu \widehat T_{\rm ref}^{0\mu}+ \frac{1}{2}\Omega_{\alpha\beta} \widehat J_{\rm ref}^{0,\alpha\beta}+\zeta \widehat J^0\right)\,.
\end{equation}
Although the total angular momentum is conserved, the orbital angular momentum is not, thus at global equilibrium $\tilde \beta_\mu$ must have zero vorticity and at local equilibrium its vorticity must be of higher order in the gradient expansion than $\Omega_{\alpha\beta}$. Separating the total angular momentum into orbital and spin parts, and combining the orbital part with the energy-momentum term, $\Omega_{\alpha\beta}$ can be recast as the leading order contribution to the thermal vorticity of the shifted velocity $\beta_\mu=\tilde \beta_\mu-\Omega_{\mu\beta}x^\beta$.
}

We transform the density operator in \eqref{eq:W0} with \eqref{eq:Wtransf2}  using the spurious currents \eqref{eq:pseudogaugeSpin} with $\widehat \Phi^{\mu,\alpha\beta}=-(\widehat S^{\mu,\alpha\beta}-\widehat S_{\rm ref}^{\mu,\alpha\beta})$ and $\zeta_\mu=\beta_\mu$. In this case we obtain the density operator in \cite{Becattini:2025twu} plus the additional terms
\begin{equation}\label{eq:deltaW}
    \delta \widehat W_{\rm ref}=\int d^d x \left(\frac{1}{2}\partial_i\beta_0\widehat S_{\rm ref}^{0,0i} +\partial_{(i}\beta_{j)}\widehat S_{\rm ref}^{i,0j}+\Omega^{(1)}_{\alpha\beta}\widehat S_{\rm ref}^{0,\alpha\beta}\right)\,.
\end{equation}
Note that $\delta \widehat W_{\rm ref}=0$ at global equilibrium, even if the vorticity is nonzero. As before, further transformations \eqref{eq:Wtransf2} using the spurious currents \eqref{eq:pseudogaugeSpin} with $\zeta_\mu=\beta_\mu$ effectively act as pseudogauge transformations \eqref{eq:pseudogauge} that leave the operator $\widehat S_{\rm ref}^{\mu,\alpha\beta}$ invariant. In this sense all  density operators \eqref{eq:W0} can be made pseudogauge invariant, with the difference between the physical states encoded in \eqref{eq:deltaW}. However, when the operator  $\widehat S_{\rm ref}^{\mu,\alpha\beta}$ is nonzero, the density operator depends on quantities that are not conserved currents.

This provides a physical understanding of why the invariant density operator is equivalent to the density operator in ``Belinfante pseudogauge.'' It must be the case in order for the local equilibrium density operator to be dependent only on conserved charges of genuine symmetries. Note that the local equilibrium operator can still be written in any pseudogauge frame, with a nonzero spin current, as long as it includes the terms \eqref{eq:DeltaWlocal}.

\section{Zilch transformations}

In addition to the pseudogauge transformations above, one can also introduce the ``Zilch'' transformation, an improvement of the spin current by a divergence
\begin{equation}\label{eq:Zilchtransf}
    \widetilde{\widehat S^{\mu,\alpha\beta}}=\widehat S^{\mu,\alpha\beta}+\partial_\sigma \widehat Z^{\sigma\mu,\alpha\beta},
\end{equation}
with $\widehat Z^{\sigma\mu,\alpha\beta}=- \widehat Z^{\mu\sigma,\alpha\beta}=- \widehat Z^{\sigma\mu,\beta\alpha}$. The associated spurious current is
\begin{equation}\label{eq:Zilch}
   \widehat  J_Z^{\mu,\alpha\beta}=\partial_\sigma \widehat Z^{\sigma\mu,\alpha\beta}\,.
\end{equation}
The Zilch improvement is implemented by a transformation \eqref{eq:Wtransf2} with the spurious current \eqref{eq:Zilch} and an antisymmetric chemical potential
\begin{equation}\label{eq:zilchchempot}
    \zeta_{ij}=\frac{1}{2}\Omega_{ij},\quad \zeta_{0i}=\zeta_{i0}=0\,.
\end{equation}
Indeed, the density operator \eqref{eq:equilvort} with an asymmetric energy-momentum tensor is transformed to\footnote{We are using a label to denote that we want to write the density operator with an asymmetric energy-momentum tensor and nonzero spin current, but note that $\widehat W_{\rm as}=\widehat W$.}
\begin{equation}
    \widetilde{\widehat W}_{\rm as}=\widehat W_{\rm as}+\int d^d x \left(\zeta_{\alpha\beta} \widehat J_Z^{0,\alpha\beta}+\partial_i \zeta_{\alpha\beta}\widehat Z^{i 0,\alpha\beta} \right)
\end{equation}
The last term in the integral vanishes, since $\zeta_{\alpha\beta}$ in \eqref{eq:zilchchempot} is constant. The first term in the integral is proportional to the vorticity and adds to the term depending on the angular momentum in \eqref{eq:equilvort}, which is modified to
\begin{equation}
\begin{split}  \widetilde{\widehat {\cal J}_{\rm as}^{ij}}=&\widehat {\cal J}_{\rm as}^{ij}+\int d^dx\, \widehat J_Z^{0,ij}\\
   &=\int d^dx\left( x^i \widehat T_{\rm as}^{0j}-x^j \widehat T_{\rm as}^{0i}+\widehat S^{0,ij}+\partial_k \widehat Z^{k 0,ij}\right)\,.
\end{split}
\end{equation}
The last two terms combine into the transformed spin current shown in \eqref{eq:Zilchtransf}.

The generalization of the Zilch transformation to local equilibrium follows similar steps to the improvement of the energy-momentum tensor: the transformation \eqref{eq:Wtransf2} is performed with the spurious current \eqref{eq:Zilch} and a chemical potential

\begin{equation}
    \zeta_{\alpha\beta}=\frac{1}{4}\left(\partial_\alpha\beta_\beta-\partial_\beta \beta_\alpha\right)\,.
\end{equation}
 The transformed operator can be written as the operator constructed in \cite{Becattini:2025twu} with the spin current given in \eqref{eq:Zilchtransf} plus terms that depend on the potential $\widehat Z^{\sigma\mu,\alpha\beta}$ and that vanish at global equilibrium
\begin{equation}
    \Delta \widehat W_Z=\int d^d x\left(\partial_{(\alpha}\beta_{\beta)} \partial_\sigma \widehat Z^{\sigma\alpha,0\beta}+ \frac{1}{2}\partial_i\partial_{[\alpha}\beta_{\beta]}\widehat Z^{i0,\alpha\beta}\right)
\end{equation}
Note that because we have used \eqref{eq:Wtransf2} to arrive at this expression, there is no actual dependence of the state on the potential of the transformation, but an analog of pseudogauge invariance emerges where the spin current and the Zilch potential are shifted simultaneously.

\section{Discussion}

In this work, we have revisited the origin of ambiguities in the construction of local thermal equilibrium density operators from pseudogauge transformations and improvement terms. We have argued that these transformations can be understood as arising from the addition to the density operator of total derivative terms depending on identically conserved currents associated with what we term spurious symmetries. Unlike genuine symmetries, they do not generate physical symmetry operations and are characterized by conserved currents that can be expressed as total divergences and therefore carry vanishing charges.

We propose that the local thermal equilibrium density operator depends only on the intensive variables conjugate to genuine currents.
This perspective naturally motivates a systematic distinction between genuine and spurious conserved currents. A natural identification in a conformal field theory is that they are separated into quasiprimary and descendant operators respectively. These identifications are extended to nonconformal theories obtained through conformal symmetry-breaking deformations.  Once this separation is made explicit, the construction of local thermal equilibrium density operators can be formulated entirely in terms of genuine conserved currents. Transformations involving currents of spurious symmetries extend from global to local equilibrium by including terms depending on derivatives of intensive variables.

As a consequence, we obtain a local equilibrium density operator that is invariant under pseudogauge transformations and, more generally, under improvement transformations of conserved currents. This resolves the residual ambiguities present in the pseudogauge-invariant construction proposed in \cite{Becattini:2025twu}. While the physical conclusions of that work remain unchanged, the present formulation places them within a more transparent and conceptually unified framework.

More broadly, the framework developed here provides an ambiguity-free characterization of local equilibrium states in relativistic quantum field theory. 
We expect these results to be particularly relevant to spin hydrodynamics---see, e.g.,~\cite{Montenegro:2026phf} for a recent pseudogauge invariant formulation---and for the interpretation of spin polarization phenomena in heavy-ion collisions. However, it should be noted that mixed states that are not exactly at local equilibrium but in a near-equilibrium state where momentum relaxation proceeds much faster than spin relaxation may be of interest to describe systems with a strong spin polarization. In the approach of  \cite{Becattini:2018duy}, this requires selecting a spin density operator and introducing a corresponding spin chemical potential independent of the thermal vorticity. Since the spin current operator is not uniquely defined, the characterization of these operators is ambiguous.

\begin{acknowledgments}
I would like to thank Francesco Becattini and Giorgio Torrieri for useful discussions and comments.

This work is partially supported by the Spanish Agencia Estatal de Investigaci\'on and Ministerio de Ciencia,
Innovaci\'on y Universidades through the Grants No. PID2021-123021NB-I00 and No. PID2024-161500NB-I00.
\end{acknowledgments}

\bibliography{refs.bib}

\end{document}